\documentclass[letterpaper, 10 pt, conference]{ieeeconf}  

\IEEEoverridecommandlockouts                              

\usepackage{times}
\usepackage{amsmath}
\usepackage{xspace}
\usepackage{float}
\usepackage{balance}
\usepackage{amssymb,latexsym,amsfonts}
\usepackage{xkeyval}
\usepackage{graphicx}
\usepackage{caption}
\usepackage{subcaption}
\usepackage{eso-pic}
\usepackage{epsfig}
\usepackage{url}
\usepackage{color}
\usepackage{tikz}
\usetikzlibrary{automata}
\usetikzlibrary{shapes,arrows}
\usepackage{pgfplots}
\usepackage{pgfplotstable}
\usepackage[latin1]{inputenc}
\usepackage{dcolumn}
\usepackage[algosection,linesnumbered,lined, ruled]{algorithm2e}
\usepackage{mathtools}
\usepackage[T1]{fontenc}
\usepackage{mathtools}
\usepackage{soul}
\usepackage{siunitx}
\usepackage{leftidx}
\usepackage{enumerate}

\usepackage{physics}
\usepackage{amsmath}
\usepackage{tikz}
\usepackage{mathdots}
\usepackage{yhmath}
\usepackage{cancel}
\usepackage{color}
\usepackage{siunitx}
\usepackage{array}
\usepackage{multirow}
\usepackage{amssymb}
\usepackage{gensymb}
\usepackage{tabularx}
\usepackage{extarrows}
\usepackage{booktabs}
\usepackage{cite}

\usetikzlibrary{fadings}
\usetikzlibrary{patterns}
\usetikzlibrary{shadows.blur}
\usetikzlibrary{shapes}

\usetikzlibrary{patterns}

\newtheorem{theorem}{Theorem}[section]
\newtheorem{definition}[theorem]{Definition}
\newtheorem{lemma}[theorem]{Lemma}
\newtheorem{proposition}[theorem]{Proposition}

\DeclareMathOperator*{\argmax}{arg\,max}

\newcommand{\prob}{\mathbb{P}}
\newcommand{\expectation}[1]{\underset{#1}{\mathbb{E}}}
\newcommand{\indicator}[2]{\mathbb{I}_{ \left\{ #1 \right\}}\!\!\left( #2 \right)}

\newcommand{\allagents}{\mathbb{D}}
\newcommand{\horizon}{L}

\newcommand{\hkfull}{{\mathrm{h}_{k}^{\allagents}}}
\newcommand{\hkr}{{\mathrm{h}_{k}^{r}}}
\newcommand{\hkrp}{{\mathrm{h}_{k}^{r'}}}
\newcommand{\chk}{{\prescript{c}{}{\mathrm{h}}_{k}^{r,r'}}}
\newcommand{\dhkrrp}{{\Delta \mathrm{h}_{k}^{r,r'}}}
\newcommand{\dhkrpr}{{\Delta \mathrm{h}_{k}^{r',r}}}
\newcommand{\dhk}{{\Delta \mathrm{h}_{k}}}

\newcommand{\Hkfull}{H_{k}^{\allagents}}
\newcommand{\Hkrp}{H_{k}^{r'}}

\newcommand{\dHkrrp}{\Delta H_{k}^{r,r'}}
\newcommand{\dHkrpr}{\Delta H_{k}^{r',r}}

\newcommand{\Hkfullspace}{\mathcal{H}_{k}^{\allagents}}
\newcommand{\tildehkfull}{{\tilde{h}_{k}^{\allagents}}}

\newcommand{\Hkrpspace}{\mathcal{H}_{k}^{r'}}
\newcommand{\tildehkrp}{{\tilde{h}_{k}^{r'}}}

\newcommand{\dHkrrpspace}{\Delta \mathcal{H}_{k}^{r,r'}}
\newcommand{\tildedhkr}{{\Delta \tilde{h}_{k}^{r}}}

\newcommand{\dHkrprspace}{\Delta \mathcal{H}_{k}^{r',r}}
\newcommand{\tildedhkrp}{{\Delta \tilde{h}_{k}^{r'}}}

\newcommand{\aoptfull}{a_{k+}^{*}}
\newcommand{\action}{a_{k+}}
\newcommand{\selectedaoptr}{\prescript{(r)}{}{a}_{k+}}
\newcommand{\selectedaoptrp}{\prescript{(r')}{}{a}_{k+}}

\newcommand{\Aopt}{A_{k+}^{*}}
\newcommand{\Aoptrp}{\prescript{(r')}{}{A}_{k+}^{*}}

\newcommand{\Actions}{\mathcal{A}^{\horizon}}

\newcommand{\Jraction}{\Delta J^{r}_{\selectedaoptr}}

\newcommand{\bkfull}{b_{0:k}^{\allagents}}
\newcommand{\bfull}[1]{b_{#1}^{\allagents}}
\newcommand{\bkc}{b_{0:k}^{c}}
\newcommand{\bkr}{b_{0:k}^{r}}
\newcommand{\bkrp}{b_{0:k}^{r'}}

\newcommand{\COMM}{\mathrm{COMM}}

\newcommand{\algorithmname}{Dec-OAC-POMDP-OL\xspace}
\newcommand{\shortalgorithmname}{DecOAC-OL\xspace}
\newcommand{\MRAC}{MRAC\xspace}
\newcommand{\MROAC}{MROAC\xspace}
\newcommand{\MLOAS}{MLOAS\xspace}
\newcommand{\OAS}{OAS\xspace}
\newcommand{\NEPG}{NEPG\xspace}

\title{\LARGE \bf
Towards Optimal Performance and Action Consistency Guarantees in Dec-POMDPs with Inconsistent Beliefs and Limited Communication
}

\author{Moshe Rafaeli Shimron$^{1}$ and Vadim Indelman$^{2}$ 
\thanks{{$^{1}$} Moshe Rafaeli Shimron is with the Technion Autonomous Systems Program (TASP),
        Technion - Israel Institute of Technology, Haifa 32000, Israel
        {\tt mosh305@campus.technion.ac.il}.}%
\thanks{{$^{2}$} Vadim Indelman is with the Stephen B. Klein Faculty of Aerospace Engineering and with the Faculty of Data and Decision Sciences,
        Technion - Israel Institute of Technology, Haifa 32000, Israel
        {\tt vadim.indelman@technion.ac.il}.}
}

\begin{document}

\maketitle
\thispagestyle{empty}
\pagestyle{empty}


\begin{abstract}\label{sec:abstract}
Multi-agent decision-making under uncertainty is fundamental for effective and safe autonomous operation.
In many real-world scenarios, each agent maintains its own belief over the environment and must plan actions accordingly.
However, most existing approaches assume that all agents have identical beliefs at planning time, implying these beliefs are conditioned on the same data. Such an assumption is often impractical due to limited communication. 
In reality, agents frequently operate with inconsistent beliefs, which can lead to poor coordination and suboptimal, potentially unsafe,   performance.
In this paper, we address this critical challenge by introducing a novel decentralized framework for  optimal joint action selection that explicitly accounts for belief inconsistencies.
Our approach provides probabilistic guarantees for both action consistency and performance with respect to open-loop multi-agent POMDP  (which assumes all data is always communicated), and selectively triggers communication only when needed.  
Furthermore, we address another key aspect of whether, given a chosen joint action, the agents should share data to improve expected performance in inference. 
Simulation results 
show our approach outperforms state-of-the-art algorithms. 


\end{abstract}
\section{Introduction}\label{sec:introduction}

The field of Autonomous Robotics, and specifically the Multi-Agent Systems (MAS), has advanced significantly in the recent years, with various applications as autonomous driving, surveillance, search and rescue, warehouse operations, and many more.
Decision making under uncertainty and partial observability are at the core of autonomous systems, and introduce great challenges for single- and multi-agent systems. These problems are often formulated as a Decentralized Partially Observable Markov Decision Process (Dec-POMDP) or Multi-Agent POMDP (MPOMDP), where the latter explicitly assumes agents constantly share their information with each other.

A common assumption in most Dec-POMDP methods is that, at the beginning of the planning session, all agents have \emph{consistent} beliefs, meaning their beliefs about the environment are conditioned on the same information. 
Such consistency is only guaranteed in MPOMDP settings, and requires prohibitively frequent and large-scale communication.
In many scenarios, such exhaustive data exchange is impractical or impossible, and in practice only partial or compressed information may be shared,
leading to inconsistent beliefs about the environment's state.
Crucially, when the agents hold inconsistent beliefs, the state-of-the-art Dec-POMDP approaches may fail to achieve proper coordination between the agents, leading to sub-optimal performance and, in some cases, even unsafe or hazardous outcomes.

\begin{figure}[t]
    \centering
    \begin{subfigure}[t]{0.3\linewidth}
        \includegraphics[width=\textwidth]{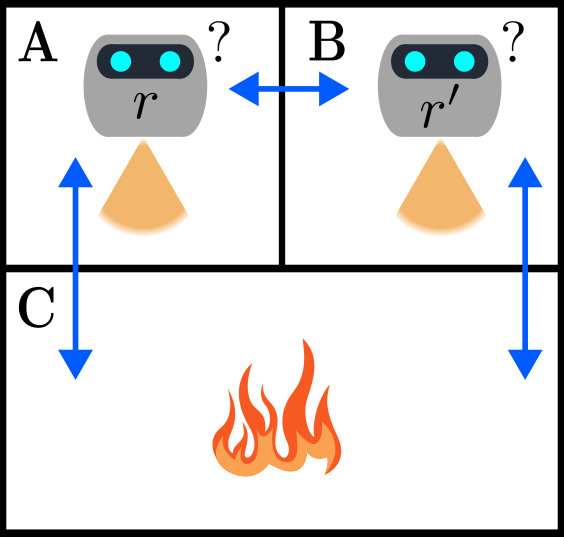}
        \caption{}
        \label{fig:motivation-example-layout}
    \end{subfigure}
    ~~~~
    \begin{subfigure}[t]{0.3\linewidth}
        \includegraphics[width=\textwidth]{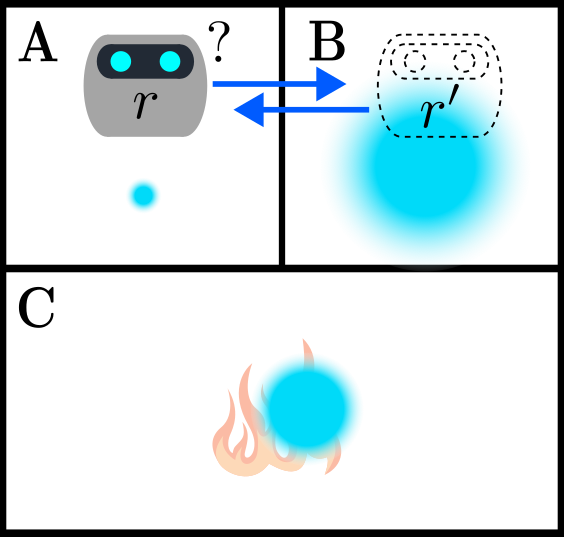}
        \caption{}
        \label{fig:motivation-example-r-selection}
    \end{subfigure}

    \begin{subfigure}[t]{0.3\linewidth}
        \includegraphics[width=\textwidth]{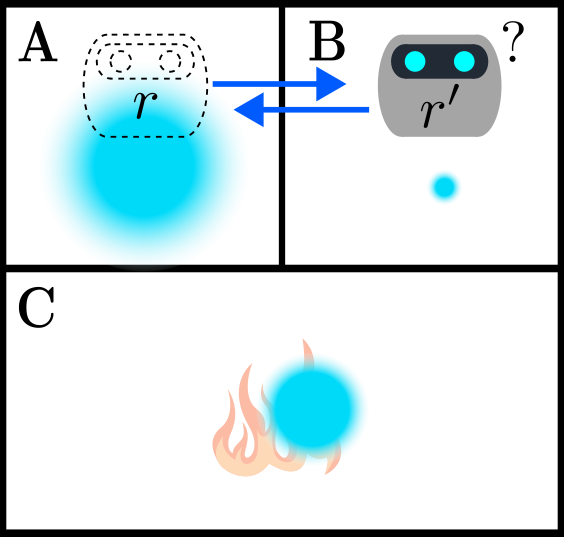}
        \caption{}
        \label{fig:motivation-example-r'-selection}
    \end{subfigure}
    ~~~~
    \begin{subfigure}[t]{0.3\linewidth}
        \includegraphics[width=\textwidth]{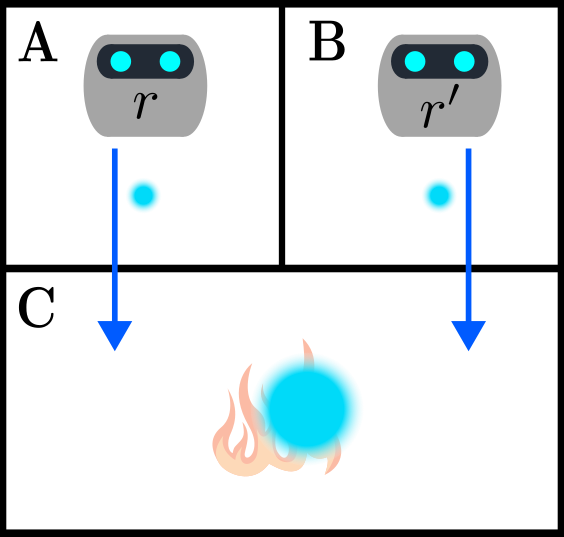}
        \caption{}
        \label{fig:motivation-example-full-history-selection}
    \end{subfigure}
    \caption{\small
        Example of two agents in a collaborative fire detection task.
        Figure~\ref{fig:motivation-example-layout} shows the initial layout and the possible movements of the agents.
        Each agent holds an observation of her current cell (in orange), which indicates with high certainty the cell is Empty.
        The agents did not share these observations, which makes the agents' beliefs inconsistent.
        Figures~\ref{fig:motivation-example-r-selection} and~\ref{fig:motivation-example-r'-selection} show the agent's inconsistent beliefs (blue circles represent the amount of uncertainty of a cell). 
        Each agent, based on the information available to her, is highly certain about her cell and highly uncertain about the other agent's cell.
        So, both agents consistently select each agent moves to the other agent's cell, thus satisfying \MRAC.
        On the other hand, Figure~\ref{fig:motivation-example-full-history-selection} shows that when considering all the data in the system, i.e.~in an MPOMDP setting, the uncertainty of cell C is higher than the uncertainties of cells A and B, thus both agents consistently select both of them to move down and observe cell C.
    }
    \label{fig:motivation-example}
\end{figure}

In this work, we address the important gap of coordinating and optimizing the performance of MAS when agents operate with inconsistent beliefs.
This is the first work, to our knowledge, that addresses the \emph{consistency and optimality of decision making altogether in Dec-POMDPs with inconsistent beliefs}.
The core ideas of our approach are (i) mimicking open-loop MPOMDP planning by each agent reasoning about other agents' information that is unavailable to her; (ii) utilizing the concept of \emph{multi-robot action consistency} (\MRAC) \cite{Kundu24iros,Kundu25arxiv} to ensure, probabilistically or deterministically, that each of the agents selects the same  joint action which is optimal with respect to an open-loop MPOMDP, i.e.~considering all data in the system; and (iii) reasoning if, \emph{given} the chosen joint action, the agents should share data to improve performance.

Consider the toy example shown in  Figure~\ref{fig:motivation-example} of a collaborative fire detection task. In decentralized POMDP planning,  
where agents have  inconsistent beliefs due to unshared observations, prior work \cite{Kundu24iros,Kundu25arxiv} allows the agents to select a consistent joint action which is optimal with respect to their \emph{individual} available information (comprising local information and data shared by other agents). Yet, when considering an MPOMDP planner, a different consistent joint action is selected, which is optimal with respect to  all information in the system and achieves better performance.

	

The main contributions of this paper are as follows:
First, we propose a novel decentralized algorithm, \algorithmname (Decentralized Optimal Action Consistent POMDP Open-Loop), that is able to detect an optimal joint action sequence with respect to open-loop MPOMDP and ensures \MRAC, both with formal (probabilistic or deterministic) guarantees, by explicitly accounting for inconsistent beliefs.
Our decentralized open-loop algorithm utilizes a distribution over the optimal joint action sequence with respect to all the information in the system for action selection, and mimics these calculations as performed by other agents to ensure a consistent action selection.
Second, we describe a mechanism for analyzing the performance gap between the selected joint action with respect to MPOMDP, and the expected performance in inference considering the available information to the agents. This distribution can be used to design strategies for improving performance by selectively triggering data sharing.
Finally, we leverage calculation reuse to improve the computational complexity of our approach in the special case of state-dependent rewards.

\section{Related Work}\label{sec:related_work}

MPODMPs and Dec-POMDPs have been actively investigated in the last  decades (see, e.g., \cite{Oliehoek16book,Pynadath02jair,Amato15aaai,Capitan13ijrr}). For instance, \cite{Amato15aaai} and \cite{Capitan13ijrr}, focus on scalability of MPOMDP settings, where the latter work shows a Dec-POMDP can be reduced to a small local MPOMDP between a subset of agents that can communicate.  These works address MPOMDPs and so they assume consistent beliefs at each planning step via communication between  agents.

Other works such as \cite{Atanasov15icra,Indelman17arj,Regev17arj}  focused on cooperative Multi-Robot Belief Space Planning (MR-BSP), in the context of cooperative active SLAM and inference. While the concepts are general, the specific approaches in these works are restricted to  high-dimensional Gaussian distributions.

Partially related to this setting are non-cooperative POMDPs, where agents have individual tasks that correspond to different reward functions. Existing works in this setting typically formulate the problem within the framework of dynamic games and reason about the Nash equilibrium of the multi-agent system  (see,  e.g., \cite{Becker24arxiv,Schwarting21tro,So23icra}). 

Yet, all the works presented so far make the prevailing assumption that the agents' beliefs at planning time are consistent, i.e.~conditioned on the same data\footnote{Nonparametric beliefs can be inconsistent also when conditioned on the same data. In this paper we do not consider such a setting.}. 
The works \cite{Wu11ai, Kundu24iros, Kundu25arxiv} are arguably the most similar to ours.
The work \cite{Wu11ai} introduce an online planning algorithm for Dec-POMDPs, explicitly considering different histories between the agents at planning and ensuring coordination.
Yet, the action selection process is based only on \emph{common} information and does not take into account the unshared information.
Moreover, the authors describe the term "inconsistent beliefs" when an agent acquires an observation that contradicts their consistent common information, which in turns triggers the agents to communicate their entire unshared information, and by doing so become consistent.

The works \cite{Kundu24iros,Kundu25arxiv} address decentralized multi-agent POMDP planning with inconsistent beliefs through the notion of \emph{action consistency} (AC).
The concept of \emph{action consistency}, described in \cite{Indelman16ral,Elimelech22ijrr,Kitanov24ijrr}, captures the observation that two decision making problems are equivalent if both of them prefer a certain action over the other actions, regardless of the objective values of the actions in each problem.
This concept was used for simplification of single-agent POMDP problem (see, e.g., \cite{Barenboim23ral,LevYehudi24aaai,Shienman22isrr,Sztyglic22iros,Zhitnikov22ai}).
In the work \cite{Kundu24iros}, each of the agents reasons about the possible beliefs of the other agents, to ensure \MRAC is satisfied by selecting the same joint action, or trigger communication to improve the consistency guarantees.
The work \cite{Kundu25arxiv} elaborately extends the latter approach to provide probabilistic \MRAC guarantees, and introduce simplifications that enable applicability to high-dimensional and continuous spaces. 
Even though \cite{Kundu24iros,Kundu25arxiv} reason about the unshared information of the other agents to ensure \MRAC, 
the action selection process itself considers only  the \emph{available information} of each agent. As a result, \cite{Kundu24iros,Kundu25arxiv}  lack optimality guarantees with respect to the \emph{full information} of the system (MPOMDP setting). 
 Additionally, these works do not discuss data sharing given a joint action to improve expected performance in inference.


\section{Preliminaries}\label{sec:preliminaries}

We focus on a
Dec-POMDP setting, defined by the tuple $ \langle \allagents, \mathcal{X} , \mathcal{A} , \mathcal{O} , b_0 , T , O , \rho \rangle $, where:
$ \allagents = \{ 1 , ... , n \} $ is the set of $n$ agents.
$ \mathcal{X} $ is the state space.
$ \mathcal{A} = \times_{ i \in \allagents } \mathcal{A}^{i} $ is the joint action space with each $\mathcal{A}^{i}$ being the individual action space of agent $i$.
$ \mathcal{O}  = \times_{ i \in \allagents } \mathcal{O}^{i} $ is the joint observation space with each $\mathcal{O}^{i}$ being the individual observation space of agent $i$.
$b_0$ is the prior belief about the state, which is known to all the agents in the system.
$ T \left( x , a , x' \right) = \prob \left( x' \mid x , a \right) $ is the transition model where $ a \in \mathcal{A}$ and $ x, x' \in \mathcal{X} $ are the joint action and states, respectively.
$ O \left( x , o \right) = \prob \left( o \mid x \right) $ is the joint observation model where $ o \in \mathcal{O} $ and $ x \in \mathcal{X} $ are the joint observation and state, respectively. The general belief-dependent reward function
$ \rho : \mathcal{B} \times \mathcal{A} \to \mathbb{R} $ represents the collaborative task of the agents in the system, with $ \mathcal{B} $ denoting the belief space.
We assume that the agent's observations are independent given the state, i.e. for a joint observation $ o = \left( o^{1} , ... , o^{n} \right) $, $ O \left( x , o \right) = \prod_{i \in \allagents} O^{i} \left( x , o^{i} \right) $ where each $ O^{i} \left( x, o^{i} \right) = \prob^i \left( o^{i} \mid x \right) $ is the observation model of agent $i$ and each $ o^{i} \in \mathcal{O}^{i} $ is the observation of agent $i$.



Generally, each agent has access to her local history, i.e.~her individual actions and acquired observations. 
When the agents are able to share \emph{all} their actions and observations,
this corresponds to an MPOMDP,
which can be reduced to a (big) POMDP problem \cite{Amato15aaai}.
In such setting, the agent's belief at each time $k$ is conditioned on $ \hkfull \triangleq \left\{ b_{0}, a_{0:k-1} , o_{1:k} \right\} $,
i.e.~the initial belief, all the joint actions and all the joint observations of the agents in the system.
We shall refer to $\hkfull$ as the true \emph{full joint history} in the system up to time $k$.

For simplicity of calculations, in this paper we consider a smoothing formulation and define the beliefs to be over all the states up to time $k$, i.e.~$ x_{0:k} \triangleq \{ x_{0} , \ldots , x_{k} \} \in \mathcal{X}^{k+1} $.
The belief of the agents in the MPOMDP setting is,
\begin{equation}\label{eq:belief-full-history}
    \bkfull \left[ x_{0:k} \right] 
    = \prob \left( x_{0:k} \middle| \hkfull \right)
    = \prob \left( x_{0:k} \middle| b_{0}, a_{0:k-1} , o_{1:k} \right).
\end{equation}
%
In the sequel, we will often use $\bkfull \left( \hkfull \right)$ to explicitly denote the history the belief is conditioned on.

Considering an open-loop setting, at each planning session the agents find the optimal joint action sequence\footnote{We use "joint action sequence" and "joint action" interchangeably. } according to the objective function defined as, 
\begin{equation}\label{eq:objective-function-open-loop}
    \begin{aligned}
        J ( \bkfull , \action )
        = \expectation{ o_{k+} \mid \bkfull , \action } \left[ \sum_{l=0}^{\horizon-1} \rho \left(\bfull{k+l} , a_{k+l} \right) \right]
        ,
    \end{aligned}
\end{equation}
where $L$ is the planning horizon, 
$\action \triangleq \{ a_{k}, ... , a_{k+\horizon-1} \} \in \Actions $ is the joint action sequence, and
$ o_{k+} \triangleq \{ o_{k+1}, ... , o_{k+\horizon-1} \} \in \mathcal{O}^{\horizon-1} $ are the future joint observations of the agents. 
Then, the optimal joint action with respect to the true full joint history is,
\begin{equation}\label{eq:optimal-joint-action-selection-open-loop}
    \aoptfull = \argmax_{ \action \in \Actions } J ( \bkfull , \action )
    .
\end{equation}
%



\subsection{Problem Formulation}\label{subsec:problem-formulation}




We assume in this work a setting where, in contrast to MPOMDP, agents have only limited capability to share their actions and observations with each other.
For simplicity, when data sharing occurs, we assume it is noise-free and instantaneous.
In such a setting, the agents do not have access to the full joint history, so their histories, and therefore beliefs, are generally not identical at planning time.
For comparison, in a Dec-POMDP setting the initial beliefs of all  agents are also assumed to be identical, and the individual (offline-learned) policies only consider local data of each of the agents, i.e.~without data sharing.
Instead, in our setting, the agents can \emph{decide} to share \emph{some} of the data with each other, and this data sharing impacts the decision-making. 
However, such a decentralized planning scheme may result in \emph{suboptimal} performance, compared to MPOMDP. 
Moreover, it may lead to \emph{inconsistent} and potentially unsafe decision-making.
We formally define these key aspects as follows.

\begin{definition}\label{def:multi-robot-action-consistency}
    \textbf{Multi-Robot Action Consistency (\MRAC).}
    Consider a multi-agent system with $n$ agents, where each agent $i$ 
    selects a joint-action $\prescript{(i)}{}{a}$.
    The system satisfies \emph{\MRAC} if and only if $\forall i \neq j: \prescript{(i)}{}{a} = \prescript{(j)}{}{a}$, i.e. all the agents select the same joint action.
\end{definition}

\begin{definition}\label{def:multi-robot-optimal-action-consistency}
    \textbf{Multi-Robot Optimal Action Consistency (\MROAC)}.
    Consider a multi-agent system with $n$ agents, where each agent $i$ 
    selects a joint-action $\prescript{(i)}{}{a}$.
    Let the joint-action $a^{*}$ be the optimal joint-action with respect to the  full joint history in the system~\eqref{eq:optimal-joint-action-selection-open-loop}.
    The system satisfies \emph{\MROAC} if and only if, $\forall i : \prescript{(i)}{}{a} = a^{*}$, i.e. all agents select the optimal joint-action with respect to the full joint history.
\end{definition}


\MROAC is a specific case of \MRAC, where the consistent joint action selected by all  agents is also the optimal joint action according to the full joint history. 
Note the optimality in \MROAC is with respect to an open-loop MPOMDP.%

In this paper we address the problem of a decentralized selection of provably \emph{optimal} and \emph{consistent} joint action, with respect to the full joint history, considering agents have access to inconsistent beliefs. 
Our proposed decentralized algorithm, \algorithmname, is described in Section \ref{sec:approach}: 
Section \ref{sec:optimal-joint-action-selection} presents
how an agent reasons about selecting an optimal joint action with respect to the full joint history,
with formal 
 deterministic and probabilistic, 
optimality guarantees.
Section \ref{sec:optimal-multi-robot-action-consistency} presents
how agents verify \MRAC for their selected joint action, also with formal guarantees.
Section \ref{sec:performance-improvement-via-communication} presents a mechanism for agents to communicate information to improve their performance, even when an optimal joint action was selected.
Section \ref{sec:calculation-reuse-in-state-dependent-rewards} shows 
how calculations can be reused to reduce computational complexity for the specific case of state-dependent rewards. 
Section \ref{sec:simulation-results} shows performance evaluation of our approach, 
compared to several state-of-the-art (SOTA) algorithms.

\section{Approach}\label{sec:approach}

In this section we describe the steps of our proposed decentralized algorithm \algorithmname for selecting a consistent and optimal joint action with formal \MROAC guarantees considering the agetns have inconsistent beliefs at planning time. We then present a strategy for improving the expected agents' performance in inference for a chosen joint action. 
Additionally, we show how we can improve the computational complexity of our approach via calculation reuse in the specific case of state-dependent rewards.


\subsection{Agents' Histories and Beliefs Definitions}\label{subsec:agents-histories-and-beliefs-definitions}

\begin{figure}[t]
    \begin{subfigure}[t]{0.32\linewidth}
        \includegraphics[width=\textwidth]{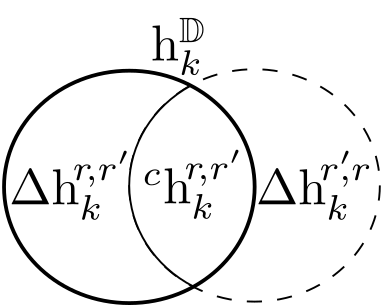}
        \caption{}
        \label{fig:local-history-r}
    \end{subfigure}
    \hfill
    \begin{subfigure}[t]{0.32\linewidth}
        \includegraphics[width=\textwidth]{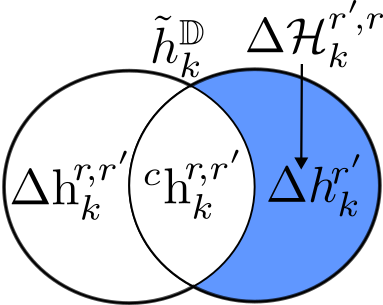}
        \caption{}
        \label{fig:possible-full-history-r}
    \end{subfigure}
    \hfill
    \begin{subfigure}[t]{0.32\linewidth}
        \includegraphics[width=\textwidth]{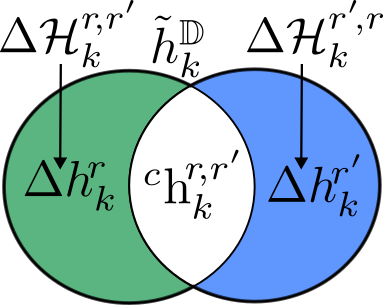}
        \caption{}
        \label{fig:possible-full-history-c}
    \end{subfigure}
    \caption{\small
        Figure~\ref{fig:local-history-r} shows the available history of agent $r$ in bold ($\chk$, $\dhkrrp$), and the available history of agent $r'$ in dashed ($\chk$, $\dhkrpr$), the full joint history $\hkfull$ is the union of them.
        Figure~\ref{fig:possible-full-history-r} shows possibilities of full joint histories $\tildehkfull$ from the perspective of agent $r$, by reasoning over the unshared data of agent $r'$, $\tildedhkrp$.
        Figure~\ref{fig:possible-full-history-c} shows possibilities of full joint histories $\tildehkfull$ by reasoning over the unshared data of agents $r$ and $r'$, $(\tildedhkrp, \tildedhkr)$.
    }
    \label{fig:histories-diagram}
\end{figure}

We consider a setting with two agents, denoted as $\allagents = \left\{ r, r' \right\}$. 
We define the histories available to agents $r$ and $r'$ as $ \hkr \subseteq \hkfull $ and $ \hkrp \subseteq \hkfull $, respectively.
These histories include the local data of the corresponding agent, as well as data shared by the other agent.
The part of the histories that is available to both agents is called the \emph{common history}, and denoted as $ \chk \triangleq \hkr \cap \hkrp $.
The part of the history that agent $r$ did not share with agent $r'$ is denoted as $ \dhkrrp \triangleq \hkr \backslash \chk $. 
Similarly, the part of the history that agent $r'$ did not share with agent $r$ is denoted as $ \dhkrpr \triangleq \hkrp \backslash \chk $.
%
 Figure~\ref{fig:local-history-r} illustrates the different histories defined above. 

The \emph{agents' beliefs} are conditioned on the histories available to them, i.e.~
$\bkr [ x_{0:k} ] \triangleq \prob \left( x_{0:k} \middle| \hkr \right) = \prob \left( x_{0:k} \middle| \chk , \dhkrrp \right) $, and $ \bkrp [ x_{0:k} ] \triangleq \prob \left( x_{0:k} \middle| \hkrp \right) = \prob \left( x_{0:k} \middle| \chk , \dhkrpr \right) $ for agents $r$ and $r'$.

\subsection{Optimal Joint Action Selection}\label{sec:optimal-joint-action-selection}

The first step of algorithm \algorithmname requires calculating an optimal joint action in~\eqref{eq:optimal-joint-action-selection-open-loop} with respect to the full joint history of the system. 
However, in our setting, each agent $r$ has access only to her own history $\hkr$. 
We consider the full joint history to be a RV, which we denote as $\Hkfull$,
taking values from the full joint history space, denoted as $\Hkfullspace$. 
This RV induces a distribution over the optimal joint action RV, denoted as $\Aopt$.
From the perspective of  agent $r$, the distribution of $\Aopt$ is,
%
\begin{equation}\label{eq:optimal-joint-action-dist-from-perspective-of-r}
    \begin{aligned}
        \prob \left( \Aopt \middle| \hkr \right)
        = \expectation{ \Hkfull \mid \hkr } \left[ \prob \left( \Aopt \middle| \hkr, \Hkfull \right) \right]
        ,
    \end{aligned}
\end{equation}
%
where the expectation over $\Hkfull$ considers all possible realizations of the full joint history $ \tildehkfull \in \Hkfullspace $.
%
%
\begin{definition}\label{def:optimal-action-guarantee}
    \textbf{Optimal Action Guarantees.}
    Given the distribution over the optimal joint action $ \prob \left( \Aopt \middle| \hkr \right) $,
    each joint action $\action$ has a \emph{Probabilistic Optimal Action guarantee} of $ \prob \left( \Aopt = \action \middle| \hkr \right) $, with respect to the true full joint history $\hkfull$.
    If there exists a joint action with probability of 1, then this joint action has a \emph{Deterministic Optimal Action guarantee}, with respect to the true full joint history $\hkfull$.
\end{definition}

\begin{proposition}
If there exists a joint action $\action$ with a Deterministic Optimal Action guarantee, then necessarily $ \action $ is the optimal joint action with respect to the true full joint history defined in \eqref{eq:optimal-joint-action-selection-open-loop}, i.e.~$\action \equiv \aoptfull$.

\begin{proof}
    Let $\action$ be a joint action with a Deterministic Optimal Action guarantee, i.e. $ \prob \left( \Aopt = \action \middle| \hkr \right) = 1 $.
    From \eqref{eq:optimal-joint-action-dist-from-perspective-of-r}, it follows that\footnote{\label{fn:drop-hkr-conditioned-on-hkfull}When conditioned on $\tildehkfull$, $\hkr$ can be dropped in $\prob \left( \Aopt \middle| \hkr, \tildehkfull \right)$.} $ \prob \left( \Aopt = \action \middle| \tildehkfull \right) = 1 $, for all realizations of the full joint history $ \tildehkfull \in \Hkfullspace $ with non-zero likelihood, i.e. $\prob \left( \Hkfull = \tildehkfull \middle| \hkr \right) \neq 0$.
    In particular, this holds for the true full joint history realization $\hkfull$, which means that\footref{fn:drop-hkr-conditioned-on-hkfull} $ \prob \left( \Aopt = \action \middle| \hkfull \right) = 1 $, and therefore necessarily $\aoptfull \equiv \action$.
\end{proof}
\end{proposition}


The distribution $ \prob \left( \Aopt \middle| \hkr \right) $  allows to design different strategies for selecting the optimal joint action.
We define the notion of an Optimal Action Selection strategy $\phi$, and consider one such strategy.
\begin{definition}\label{def:optimal-action-selection-strategy}
    \textbf{Optimal Action Selection (\OAS) Strategy.}
    An operator $\phi$ defines an \emph{Optimal Action Selection} strategy if it 
    gets the distribution of the optimal joint action as input, $ \prob \left( \Aopt \middle| \cdot \right) $, and outputs a selected optimal joint action, or a communication trigger. 
    Formally, $\phi :  \Delta \left( \Aopt \right)  \to \left\{ \Actions , \COMM \right\}$,
    where $  \Delta \left( \Aopt \right) $ denotes the space of all distributions of the optimal joint action RV $\Aopt$.
\end{definition}

\begin{definition}\label{def:epsilon-maximum-likelihood-optimal-action-selection}
    \textbf{Maximum Likelihood Optimal Action Selection ($\epsilon$-\MLOAS) Strategy.
    }
    Let $ \epsilon \in \left[ 0 , 1 \right] $ be a user pre-defined threshold parameter.
    The $\epsilon$-\MLOAS
    strategy selects the joint action with the highest Optimal Action probability, i.e.~$ \selectedaoptr = \argmax_{ \action \in \Actions} \prob \left( \Aopt = \action \middle| \hkr \right) $,  if its probability is above the threshold $1-\epsilon$, otherwise the strategy triggers communication.
\end{definition}


\subsubsection*{Calculating the Optimal Joint Action Distribution}\label{subsec:calculating-optimal-joint-action-distribution}


We now specify how agent $r$ can calculate the optimal joint action distribution \eqref{eq:optimal-joint-action-dist-from-perspective-of-r} in practice. 
The optimal joint action RV depends on the realizations of the full joint history.
Recalling the history definitions from Section~\ref{subsec:agents-histories-and-beliefs-definitions}, 
the true full joint history can be rewritten as, $ \hkfull = \chk \cup \dhkrrp \cup \dhkrpr $.
From the perspective of agent $r$, the unknown part of the full joint history is the unshared data of agent $r'$, i.e. $\dhkrpr$, as seen in Figure~\ref{fig:local-history-r}.
Similarly, we define a RV of the unshared data of agent $r'$ as $\dHkrpr$,  taking values from the corresponding space $\dHkrprspace$. 


Given the available history of agent $r$, the realizations of the full joint history are induced by the realizations $\tildedhkrp \in \dHkrprspace$, i.e. $\tildehkfull = \hkr \cup \tildedhkrp $, as seen in Figure~\ref{fig:possible-full-history-r}.
From each such realization, agent $r$ constructs the belief,
$
        \bkfull \left( \tildehkfull \right) \left[ x_{0:k} \right]
        = \prob \left( x_{0:k} \middle| \tildehkfull \right) 
        = \prob \left( x_{0:k} \middle| \hkr , \tildedhkrp \right),
$
and calculates the corresponding optimal joint action,
\begin{align}
    a^{*} \!\! \left( \hkr, \tildedhkrp \right)
    \!\!=\! a^{*} \!\! \left( \tildehkfull \right)
    \!\!=\! \argmax_{ \action \in \Actions } J \! \left( \bkfull  \left( \tildehkfull \right) , \action \right)
    .
    \label{eq:optimal-joint-action-from-perspective-of-r}
\end{align}
Finally, from the perspective of agent $r$, considering all realizations $\tildedhkrp \in \dHkrprspace$ of $\dHkrpr$, the likelihood of joint action $\action$ being the optimal joint action is,
\begin{equation}\label{eq:probability-of-optimal-joint-action-from-perspective-of-r}
        \prob \! \left( \Aopt \!\!=\! \action \middle| \hkr \right) 
        \!= \!\!\!\!\! \expectation{ \dHkrpr \mid \hkr } \!\! \left[ \indicator{ a^{*} \left( \hkr, \dHkrpr \right) = \action }{\!\dHkrpr\!} \! \right]
        .
\end{equation}
The computational complexity of this calculation is
$ O \big( \abs{\dHkrprspace} \cdot \abs{\Actions} \cdot \abs{\mathcal{X}}^{L} \cdot \abs{\mathcal{O}}^{L-1} \big)$.

\subsection{Multi-Robot Action Consistency for Optimal Joint Action}\label{sec:optimal-multi-robot-action-consistency}

In this section we present the second step of algorithm \algorithmname, providing formal (deterministic or probabilistic) guarantees on \MRAC.
We recall that the optimal joint action in \MROAC is defined with respect to the true full joint history (Definition~\ref{def:multi-robot-optimal-action-consistency}).
Thus far, based on Section \ref{sec:optimal-joint-action-selection} and some \OAS strategy $\phi$, agent $r$ calculated a joint action $\selectedaoptr$ which is optimal with probability $\prob  \left( \Aopt = \selectedaoptr \middle| \hkr \right)$ from  \eqref{eq:probability-of-optimal-joint-action-from-perspective-of-r}.
However, at this point, agent $r$ has no guarantee that agent $r'$ will select the same joint action. 
We present a method to calculate the probability of \MRAC for joint action $\selectedaoptr$, i.e.~the probability that agent $r'$ will also choose $\selectedaoptr$, given the same \OAS strategy $\phi$. 

%

Agent $r'$ will select a joint action according to the \OAS strategy $\phi$ and her available history $\hkrp$.
Since agent $r$ does not have access to $\hkrp$, we define it as a RV, denoted as $\Hkrp$, which takes values from the corresponding available history space of agent $r'$, denoted as $\Hkrpspace$.
From the perspective of agent $r$, the selected joint action by agent $r'$ is also a RV, denoted by $\Aoptrp$, which is induced by $\Hkrp$, 
\begin{equation}\label{eq:probability-action-selection-r-prime}
    \begin{aligned}
        \prob \left( \Aoptrp \middle| \hkr \right)
        = \expectation{ \Hkrp \mid \hkr } \left[ \prob \left( \Aoptrp \middle| \hkr, \Hkrp \right) \right]
        .
    \end{aligned}
\end{equation}
$ \prob \left( \Aoptrp = \selectedaoptr \middle| \hkr \right) $ represents the  probability, from the perspective of agent $r$, for agent $r'$ to select the same action $\selectedaoptr$.
In other words, it represents the \MRAC probability  for the selected joint action $\selectedaoptr$.
Therefore, if $ \prob \left( \Aoptrp = \selectedaoptr \middle| \hkr \right) = 1$, \MRAC is guaranteed deterministically, and otherwise it is guaranteed probabilistically.
We note that Optimal Action guarantees (Definitions~\ref{def:optimal-action-guarantee}) and \MRAC guarantees
are independent, and can be satisfied separately in different combinations.
Together, these guarantees indicate the probability of \MROAC (Definition~\ref{def:multi-robot-optimal-action-consistency}).


\begin{lemma}\label{def:probability-of-multi-robot-optimal-action-consistency}
    Let $\phi$ be an \OAS strategy, and let $\selectedaoptr$ be the joint action selected by agent $r$ according to that strategy.
    From the perspective of agent $r$, \MROAC is guaranteed \emph{probabilistically} for that joint action $\selectedaoptr$ with the probability,
    \begin{equation}\label{eq:probability-of-multi-robot-optimal-action-consistency}
    	\begin{aligned}
    		& \prob \left( \mathrm{\MROAC} \middle| \hkr , \selectedaoptr \right) = \\
    		& \quad \prob \left( \Aopt = \selectedaoptr \middle| \hkr \right)
    		\cdot
    		\prob \left( \Aoptrp = \selectedaoptr \middle| \hkr \right)
    		.
    	\end{aligned}
    \end{equation}
   %
    When both probabilities in \eqref{eq:probability-of-multi-robot-optimal-action-consistency} are equal to 1, \MROAC is guaranteed \emph{deterministically} with $\selectedaoptr  \equiv \selectedaoptrp \equiv \aoptfull$.
\end{lemma}

\subsubsection*{Calculating  $r'$ Action Selection Distribution}\label{subsec:calculating-r-prime-action-selection-distribution}

We now specify how agent $r$ can calculate \eqref{eq:probability-action-selection-r-prime}, the distribution of action selection by agent $r'$,  in practice. 
Agent $r$ is aware that agent $r'$, finds the distribution of the optimal joint  action with respect to her \emph{true} $\hkrp$, and then decides about the joint action $\selectedaoptrp$ according to the \OAS strategy $\phi$.
Agent $r'$ does so by reasoning about the unshared data of agent $r$, which corresponds to the RV $\dHkrrp$ that takes values from the space $\dHkrrpspace$.
Since the true $\hkrp$ is unknown to agent $r$, agent $r$ mimics this process for each possible realization $\tildehkrp \in \Hkrpspace$.
Similar to Section~\ref{sec:optimal-joint-action-selection}, this corresponds to reasoning about the RV $\dHkrpr$ that represents the unshared history of agent $r'$, i.e.~the part in $\hkrp$ that agent $r'$ did not share with agent $r$.


For each realization $\tildedhkrp \in \dHkrprspace$, agent $r$ constructs a realization of the history available to agent $r'$, $\tildehkrp = \chk \cup \tildedhkrp$. 
Then, for each realization $\tildedhkr \in \dHkrrpspace$, agent $r$ constructs a realization of a full joint history, $\tildehkfull = \tildehkrp \cup \tildedhkr = \chk \cup \tildedhkrp \cup \tildedhkr $, as seen in Figure~\ref{fig:possible-full-history-c}.
Importantly, one of the realizations $\tildehkfull$ \emph{necessarily} corresponds to the true value, $\hkfull$. 
The belief, conditioned on the realization of the full joint history, $\tildehkfull$, is
\begin{equation}\label{eq:joint-history-belief-of-rprime-from-perspective-of-r}
        \bkfull \left( \tildehkfull \right) \left[ x_{0:k} \right]
        = \prob \left( x_{0:k} \middle| \tildehkfull \right)
        = \prob \left( x_{0:k} \middle| \tildehkrp , \tildedhkr \right),
\end{equation}
%
%
and the corresponding optimal joint action 
is,
\begin{equation}\label{eq:optimal-joint-action-of-rprime-from-perspective-of-r}
        a^{*} \!\! \left( \tildehkrp , \tildedhkr \right)
        \! = \! a^{*} \!\! \left( \tildehkfull \right)
        \!\! = \! \argmax_{ \action \in \Actions} J \! \left( \bkfull \! \left( \tildehkfull \right) \! , \action \! \right)
        .
\end{equation}
Now, for any realization $\tildehkrp$, the \OAS strategy $\phi$ specifies a joint action $\selectedaoptrp$ according to the distribution $\prob \left( \Aopt \middle| \tildehkrp \right)$.
The latter can be calculated, similar to \eqref{eq:probability-of-optimal-joint-action-from-perspective-of-r},
\begin{equation}
    \begin{aligned}
        \prob & \left( \Aopt = \action \middle| \tildehkrp \right) \\
        & = \expectation{ \dHkrrp \mid \tildehkrp } \left[ \indicator{ a^{*} \left( \tildehkrp, \dHkrrp \right) = \action }{\dHkrrp} \right]
        .
    \end{aligned}
\end{equation}
Finally, from the perspective of agent $r$, considering all realizations of $\tildehkrp = \chk \cup \tildedhkrp $, where $\tildedhkrp \in \dHkrprspace$, the probability that agent $r'$ will select a joint action $\action$ is,
\begin{equation}\label{eq:probability-of-optimal-joint-action-selection-of-rprime-from-perspective-of-r}
    \begin{aligned}
        & \prob \left( \Aoptrp = \action \middle| \hkr \right) \\
        & = \!\!\! \expectation{ \dHkrpr \mid \hkr } \! \left[ \indicator{ \phi \left( \Aopt \middle| \chk, \dHkrpr \right) = \action }{ \dHkrpr } \right]
        .
    \end{aligned}
\end{equation}
%
The computational complexity of this calculation is
$ O \big( \abs{\dHkrprspace} \cdot \abs{\dHkrrpspace} \cdot \abs{\mathcal{A}}^{L} \cdot \abs{\mathcal{X}}^{L} \cdot \abs{\mathcal{O}}^{L-1} \big) $.



\subsection{Performance Improvement via Communication}\label{sec:performance-improvement-via-communication}

Thus far we presented a decentralized approach that enables a group of agents with inconsistent beliefs to calculate a consistent and optimal joint action with respect to the full joint history of the system, i.e.~\MROAC, with formal probabilistic guarantees \eqref{eq:probability-of-multi-robot-optimal-action-consistency}. This approach may require some communication between the agents, e.g.~as stated by the $\epsilon$-\MLOAS strategy (Definition \ref{def:epsilon-maximum-likelihood-optimal-action-selection}). However, upon declaring that \MROAC is satisfied, the agents generally have not exchanged all data with each other, and therefore still possess inconsistent histories and beliefs. 
This discrepancy raises a key question - should the agents communicate, given the same (optimal) joint action is guaranteed?

In practice, in a decentralized setting, there is a \emph{gap} between the performance that was calculated in planning and the expected performance in execution. The former is based on reasoning about the full joint histories, while the latter is based only on the available histories (beliefs) to each of the agents.
In this section we present the final step of algorithm \algorithmname, analyzing the performance gap and show that, based on it, one can design strategies that invoke communication to improve performance in execution.


Specifically, from the perspective of agent $r$, the performance in planning of the joint action sequence $\selectedaoptr$, selected by some \OAS strategy $\phi$, is calculated by the Objective Function \eqref{eq:objective-function-open-loop}, for different realizations of the full joint history $\tildehkfull \in \Hkfullspace$ (Section \ref{sec:optimal-joint-action-selection}).
However, in an online setting, usually only the first $M$ joint actions from the selected joint action sequence are executed, after which the agents replan.
The expected performance in this setting, from the perspective of agent $r$, is obtained by evaluating the rewards from the first $M$ joint actions,
\begin{equation*}\label{eq:objective-function-local-history-r-replanning}
        J^{M} \left( \bkr , \! \selectedaoptr \right)
        \! \triangleq \!\!\!\!\! \expectation{ o_{k+} \mid \bkr , \selectedaoptr } \!\! \left[ \sum_{m=0}^{M-1} \! \rho ( b^{r}_{k+m} , \prescript{(r)}{}{a}_{k+m} ) \right]
        ,
\end{equation*}
given the belief $\bkr$, that is conditioned on $\hkr$.

The performance gap between  a specific realization of the full joint history, $\tildehkfull = \hkr \cup \tildedhkrp$, and the available history of agent $r$, $\hkr$, is given by,
\begin{equation*}\label{eq:performance-gap-from-perspective-of-r-per-realization}
    \begin{aligned}
        \Delta & J^{M} \left( \hkr , \tildedhkrp , \selectedaoptr \right) \\
        & \triangleq J^{M} \left( \bkfull \left( \tildehkfull \right) , \selectedaoptr \right) - J^{M} \left( \bkr \left( \hkr \right) , \selectedaoptr \right)
        .
    \end{aligned}
\end{equation*}
Since $\dHkrpr$ is a RV from the perspective of agent $r$,  the performance gap of the selected joint action is also a RV, denoted as $ \Jraction $. Its distribution is induced by the realizations of $ \tildedhkrp \in \dHkrprspace $, 
\begin{equation}\label{eq:performance_gap_from_perspective_of_r}
    \begin{aligned}
        \prob & \left( \Jraction = y \middle| \hkr, \selectedaoptr \right) \\
        & = \!\! \expectation{ \dHkrpr \mid \hkr } \left[ \indicator{ \Delta J \left( \hkr, \dHkrpr, \selectedaoptr \right) = y }{\dHkrpr} \right]
        .
    \end{aligned}
\end{equation}
This distribution 
indicates the \emph{importance of unshared data} from other agents.
If the distribution is narrow and centered near zero, the performance of the joint action $\selectedaoptr$ is close to the expected performance at planning, so unshared data is not important. 
However, if the distribution is wide or multi-modal, this suggests a potentially large performance gap, indicating that agent $r$ may benefit from receiving unshared data from other agents to improve execution performance.

Given these insights, one can design different strategies to decide when to communicate data between agents based on the distribution \eqref{eq:performance_gap_from_perspective_of_r}. A general such strategy is formally defined, similar to \OAS, as  $\varphi :  \Delta \left( \Jraction \right)  \to \left\{ \COMM, \neg\COMM \right\}$,
where 
$\Delta \left( \Jraction \right) $ denotes the space of all distributions over the performance gap $\Jraction$.
We now consider one such strategy.
\begin{definition}\label{def:expected-performance-gap-strategy}
    \textbf{Normalized Expected Performance Gap  ($\delta$-\NEPG)  Strategy.}
    Let $\delta$ be a user pre-defined threshold parameter.
    The $\delta$-\NEPG strategy will trigger communication if and only if the normalized expected performance gap is below $\delta$, i.e. $ \expectation{} \left[ | \Jraction | \right] /  J^{M} ( \bkr , \selectedaoptr )   < \delta $.
\end{definition}

We note that these calculations can be performed as part of the calculation of the \OAS step described in~\ref{sec:optimal-joint-action-selection}, without increasing its computational complexity.




\subsection{Calculation Reuse in State-Dependent Rewards}\label{sec:calculation-reuse-in-state-dependent-rewards}

Our approach, as presented so far in Sections~\ref{sec:optimal-joint-action-selection} and \ref{sec:optimal-multi-robot-action-consistency}, requires the agents to reason about possibilities of the full joint history, and for each realization to evaluate the Objective Function to find the optimal joint action.
This process can be computationally expensive, especially when the state space, action space, or horizon are large.

In this section we analyze the special case of a \emph{state-dependent} reward function, $ \rho ( b_k , a_k) = \expectation{ x_k | b_k } \left[ R ( x_k , a_k ) \right] $, and show how, in this case,  
the above calculations can be reused for all agents, regardless of their beliefs or histories. To this end, we introduce the following Lemma.


\begin{lemma}\label{lemma:StateDepReUse}
    For state-dependent reward functions, the Objective Function \eqref{eq:objective-function-open-loop} can be expressed, for any realization $\tildehkfull$ and the corresponding belief $\bkfull \left( \tildehkfull \right)$, as,
    \begin{equation}\label{eq:objective-function-open-loop-state-dependent}
        \begin{aligned} 
            J \! \left( \bkfull\left( \tildehkfull \right) , \action \right)
            \!\! = \! \eta^{-1}  \!\!\! \expectation{ x_{0:k} \mid \bkc } \!\! \left[ \prob \! \left( \! \dhk \middle| x_{0:k} \right) \! \cdot \! g \! \left( x_{k} , \action \right) \right] \!
            ,
        \end{aligned}
    \end{equation}
    where $\bkc$ is the belief that is conditioned only on the common history $\chk$, $\dhk \triangleq \dhkrrp \cup \dhkrpr$ is the unshared data such that $\tildehkfull = \chk \cup \Delta h_k$, and  $\eta = {\prob \left( \dhk \middle| \chk \right)}$
    is the normalizer. 
    The function $ g \left( x_{k}, \action \right)$, defined as, 
    \begin{equation}\label{eq:g-fixed-state-objective-function}
        \begin{aligned}
            g \left( x_{k} , \action \right)
            \triangleq
            \expectation{ x_{k+1:k+L-1} | x_{k} , \action } \!\! \left[ \sum_{l=0}^{L-1} R ( x_{k+l} , a_{k+l} ) \right]
            ,
        \end{aligned}
    \end{equation}
    represents the value of the expected cumulative rewards for a joint action sequence $\action$ from an initial state $x_{k}$.
\end{lemma}
The proof of Lemma \ref{lemma:StateDepReUse} follows from straightforward applications of Bayes and chain rules.

We note that the function $ g \left( x_{k} , \action \right) $ is independent of the history or the future observations, and $\eta$ is
not a function of candidate actions
(i.e. can be discarded in the $\argmax$ \eqref{eq:optimal-joint-action-selection-open-loop}).

By utilizing the form of the Objective Function in \eqref{eq:objective-function-open-loop-state-dependent} at the \OAS calculations~\eqref{eq:optimal-joint-action-from-perspective-of-r} and at the \MRAC calculations~\eqref{eq:optimal-joint-action-of-rprime-from-perspective-of-r},
we can calculate the value of $ g ( x_{k} , a_{k+} ) $ \emph{only once} for each state $ x_{k} \in \mathcal{X} $ and joint action $ a_{k+} \in \Actions $, and \emph{reuse} the result for different realizations of the full joint history $\tildehkfull \in \Hkfullspace$.
This can be used in discrete cases directly (by going over all the possibilities in the expectation), and also by estimators in high-dimensional discrete cases or continuous cases, where the expectation is approximated empirically using state samples. 
The same samples of the common belief $\bkc$ can be used for different realizations $\tildehkfull \in \Hkfullspace$, and so the calculations of the function $g$ of these samples can be reused to significantly reduce the computational complexity of the planning process.
With calculation reuse, the computational complexity of the \OAS calculations (Section~\ref{sec:optimal-joint-action-selection}) is reduced to $ O \big( \abs{\dHkrprspace} \cdot \abs{\mathcal{A}}^{L} \cdot \abs{\mathcal{X}} \big) $, and the computational complexity of the \MRAC calculations  (Section~\ref{sec:optimal-multi-robot-action-consistency}) is reduced to $ O \big( \abs{\dHkrprspace} \cdot \abs{\dHkrrpspace} \cdot \abs{\mathcal{A}}^{L} \cdot \abs{\mathcal{X}} \big) $.

\section{Simulation Results}\label{sec:simulation-results}

In this section we evaluate our approach in simulations considering a collaborative fire detection scenario. We compare our algorithm \algorithmname,  using $\epsilon$-\MLOAS and $\delta$-\NEPG strategies (denoted  as \textsc{\shortalgorithmname-$\epsilon$-$\delta$} for short), with the following planning algorithms:
(1) Open loop MPOMDP planner (\textsc{MPOMDP-OL}), in which all agents have access to all the information in the system; 
(2) Open loop Dec-POMDP planner \emph{without communication capabilities} (\textsc{DecPOMDP-OL}), where agents plan only according to their available information; 
(3) \textsc{RVerifyAC-$\epsilon$}~\cite{Kundu25arxiv}, a decentralized open loop algorithm that considers inconsistent beliefs in planning and ensures \MRAC with probabilistic guarantees.
%

\begin{table}[tb]
	\centering

	\begin{tabular}{|l|c|c|c|}
		\hline
		\multirow{2}{*}{Algorithm}	& \multicolumn{3}{c|}{ Final Return } \\
									\cline{2-4}
									& Agent 1  & Agent 2  & Centralized \\
		\hline
		\textsc{\!MPOMDP-OL}  & \textbf{-1.31 $\!\!\pm\!\!$ 0.16}  & \textbf{-1.31 $\!\!\pm\!\!$ 0.16}  & \textbf{-1.31 $\!\!\pm\!\!$ 0.16} \\
		\textsc{\!DecPOMDP-OL}  & -1.53 $\!\!\pm\!\!$ 0.17  & -1.55 $\!\!\pm\!\!$ 0.18  & -1.37 $\!\!\pm\!\!$ 0.12 \\
		\textsc{\!RVerifyAC-0.3}  & -1.53 $\!\!\pm\!\!$ 0.17  & -1.55 $\!\!\pm\!\!$ 0.18  & -1.37 $\!\!\pm\!\!$ 0.12 \\
		\textsc{\!\shortalgorithmname-0.3-0.15}  & -1.55 $\!\!\pm\!\!$ 0.16  & -1.56 $\!\!\pm\!\!$ 0.17  & \textbf{-1.31 $\!\!\pm\!\!$ 0.16} \\
		\textsc{\!\shortalgorithmname-0.3-0.05}  & \textbf{-1.36 $\!\!\pm\!\!$ 0.16}  & \textbf{-1.37 $\!\!\pm\!\!$ 0.17}  & \textbf{-1.31 $\!\!\pm\!\!$ 0.16} \\
		\hline
	\end{tabular}

	\caption{\small
		Statistical results over 25 runs with different seeds, for a single planning session for the 2x2 grid, and 1 unshared observation with value Empty for each agent.
	}
	\label{tab:simulation-results-summary-1step-00observations}
\end{table}

\definecolor{greenc}{RGB}{44,160,44}    
\definecolor{redc}{RGB}{214,39,40}    

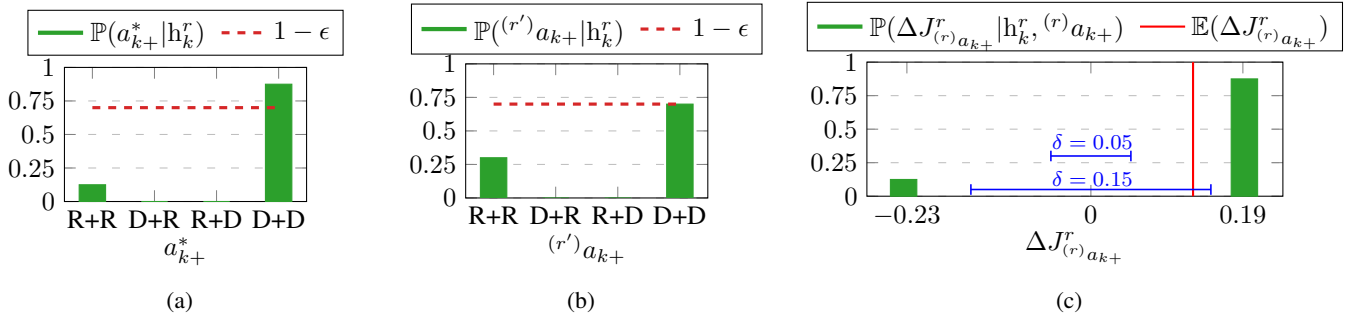
\begin{figure*}[t]
\centering

\begin{subfigure}[b]{0.27\textwidth}
	\begin{tikzpicture}
		\begin{axis}[
			legend style={at={(0.5, 1.43)}, anchor=north, legend columns=2},
			xlabel={$\aoptfull$},
			xlabel style={at={(axis description cs:0.5,0.1)},anchor=north}, 
			ymin=0, ymax=1.0,
			xtick=data,
			symbolic x coords={R+R,D+R,R+D,D+D},
			enlarge x limits=0.15,
			ytick={ 0.0, 0.25, 0.5, 0.75, 1.0},
			ymajorgrids=true,
			grid style=loosely dashed,
			height=0.14\textheight,
			width=\linewidth,
		]
			\addplot[greenc, ybar, mark=none, fill=greenc, forget plot] coordinates {(R+R,0.125) (D+R,0.0) (R+D,0.0) (D+D,0.875)};

			\addplot[redc,dashed,very thick, forget plot] coordinates {(R+R,0.7) (D+D,0.7)};


			\addlegendimage{solid, ultra thick, greenc} 
			\addlegendentry{ $\prob ( \aoptfull | \hkr )~$ }
			\addlegendimage{dashed, very thick, redc} 
			\addlegendentry{ $1-\epsilon$ }

		\end{axis}
	\end{tikzpicture}
	\caption{  }
	\label{plot:mloas-opt-action-dist-reapet1-1step-00observations}
\end{subfigure}
\hfill
\begin{subfigure}[b]{0.27\textwidth}
	\begin{tikzpicture}
		\begin{axis}[
			legend style={at={(0.5, 1.45)}, anchor=north, legend columns=2},
			xlabel={$ \selectedaoptrp $},
			xlabel style={at={(axis description cs:0.5,0.1)},anchor=north}, 
			ymin=0, ymax=1.0,
			xtick=data,
			symbolic x coords={R+R,D+R,R+D,D+D},
			enlarge x limits=0.15,
			ytick={ 0.0, 0.25, 0.5, 0.75, 1.0},
			ymajorgrids=true,
			grid style=loosely dashed,
			height=0.14\textheight,
			width=\linewidth,
		]
			\addplot[greenc, ybar, mark=none, fill=greenc, forget plot] coordinates {(R+R,0.3) (D+R,0.0) (R+D,0.0) (D+D,0.7)};

			\addplot[redc,dashed,very thick, forget plot] coordinates {(R+R,0.7) (D+D,0.7)};

			\addlegendimage{solid, ultra thick, greenc} 
			\addlegendentry{ $\prob ( \selectedaoptrp | \hkr )~$ }
			\addlegendimage{dashed, very thick, redc} 
			\addlegendentry{ $1-\epsilon$ }

		\end{axis}
	\end{tikzpicture}
	\caption{ }
	\label{plot:mloas-mrac-dist-reapet1-1step-00observations}
\end{subfigure}
\hfill
\begin{subfigure}[b]{0.4\textwidth}
	\begin{tikzpicture}
		\begin{axis}[
			legend style={at={(0.5, 1.45)}, anchor=north, legend columns=2},
			xlabel={$\Jraction$},
			xlabel style={at={(axis description cs:0.5,0.1)},anchor=north}, 
			xmin=-0.28, xmax=0.24,
			ymin=0.0, ymax=1.0,
			xtick={-0.23, 0.0, 0.19},
			ytick={ 0.0, 0.25, 0.5, 0.75, 1.0},
			ymajorgrids=true,
			grid style=loosely dashed,
			height=0.14\textheight,
			width=\linewidth,
		]

			\addplot[red, solid, thick, mark=none, forget plot] coordinates { (0.1277,0.0) (0.1277,1.0)};

			\addplot[greenc, ybar, mark=none, fill=greenc, forget plot] coordinates {(-0.2341,0.125) (0.1919,0.875)};
			

			\addplot[blue, solid, mark=|, semithick, forget plot] coordinates {(-0.15,0.05) (0.15,0.05)};
			\node[blue,font=\footnotesize,anchor=south west] at (axis cs:-0.06,0.01) {$\delta=0.15$};
			
			\addplot[blue, solid, mark=|, semithick, forget plot] coordinates {(-0.05,0.3) (0.05,0.3)};
			\node[blue,font=\footnotesize,anchor=south west] at (axis cs:-0.06,0.28) {$\delta=0.05$};

			\addlegendimage{solid, ultra thick, greenc} 
			\addlegendentry{ $\prob (\Jraction | \hkr, \selectedaoptr)~$ }
			\addlegendimage{solid, thick, color=red} 
			\addlegendentry{ $\expectation{}( \Jraction )$ }
		\end{axis}
	\end{tikzpicture}

	\caption{  }
	\label{plot:mloas-performance-gap-reapet1-1step-00observations}
\end{subfigure}
\caption{\small 
	Illustration of \algorithmname from a specific run in a 2x2 grid scenario.
	Figure~\ref{plot:mloas-opt-action-dist-reapet1-1step-00observations} shows the $\epsilon$-\MLOAS strategy with $\epsilon=0.3$, where action (D+D) is the selected optimal joint action.
	Figure~\ref{plot:mloas-mrac-dist-reapet1-1step-00observations} shows the distribution  of action selection by $r'$ in the calculations of \MRAC guarantee $\epsilon=0.3$.
	Figure~\ref{plot:mloas-performance-gap-reapet1-1step-00observations} shows the performance gap distribution of joint action (D+D) in the $\delta$-\NEPG strategy, where the red line is the normalized expected performance gap. For $\delta=0.15$ the gap is withing the threshold, and for $\delta=0.05$ the gap is beyond the threshold, which triggers communication.
}
\label{fig:mloas-planning-reapet1-1step-00observations}
\end{figure*}

\subsection{Simulation Setup}\label{subsec:simulation-setup}
The fire detection scenario consists of a grid of binary cells (Empty or Fire), with two agents located at the top left corner of the grid. 
The agents aim to reduce their uncertainty about the fire locations, maintaining a belief over Bernoulli variables representing the cells, considered statistically independent for simplicity.
To that end, the task is modeled by a negative entropy as the reward function.
We assume a deterministic transition model, and that agents know each other's actions and locations.
The observation models are stochastic, with accuracy of 0.75.

To illustrate the main contributions of our approach, both agents are initialized with the same prior belief $b_0$, but at planning time, each agent has unshared observations, such that their beliefs are inconsistent.
We perform simulations on 2 scenarios: a small 2x2 grid with fires at the bottom cells, where each agent has 1 unshared observation and performs a single planning session; and a 4x4 grid with fires at the center cells, where each agent has 2 unshared observations and performs 4 planning sessions.
Simulations were performed on a workstation equipped with an Intel Core i7-8750H CPU. 
All the algorithms were implemented in \textsc{Julia}.

\subsection{Results}\label{subsec:results}


 Table~\ref{tab:simulation-results-summary-1step-00observations} summarizes statistical performance comparison of the small 2x2 grid, where the values of the unshared observations of both agents are "Empty". 
In this case, each agent can go $\mathrm{Right}$ (denoted as R) or $\mathrm{Down}$ (denoted as D), with 4 possible joint actions (R+R), (R+D), (D+R), (D+D).
The Agent Final Return is the reward at inference when considering only the agent's available information, and the Centralized Final Return is the reward in inference when considering the full joint history.

First, we analyze in detail the joint action selection of the algorithms, aiming to demonstrate a situation where \textsc{MPOMDP-OL} and \textsc{\shortalgorithmname} choose the same action, which is different than the one chosen by \textsc{RVerifyAC} and \textsc{DecPOMDP-OL}.
In this scenario, the agents need to decide which cell (the Right cell, or the Down cell) they want to observe next, to reduce their uncertainty of it.
Considering each agent knows the prior belief and only has a single unshared observation of the Right cell, the entropy of the Right cell is still \emph{higher} than the entropy of the Down cell (due to the prior belief about the bottom cells).
Therefore, algorithms \textsc{DecPOMDP-OL} and \textsc{RVerifyAC} (with $\epsilon\!=\!0.3$), which consider only the agent's available information, select the joint action (R+R).
When considering both unshared observations of the Right cell, the entropy of the Right cell becomes \emph{smaller} than the entropy of the Down cell, despite the high prior belief about the bottom cells.
Therefore, algorithm \textsc{MPOMDP-OL} selects the joint action (D+D) instead of (R+R).
Our algorithm \textsc{\shortalgorithmname} with $\epsilon\!=\!0.3$, considering all possible values of the unshared observations of the Right cell, 
selects the joint action (D+D) with probabilistic \OAS and \MRAC guarantees, similarly to \textsc{MPOMDP-OL}.
The details of the \OAS and \MRAC guarantees calculations are presented in Figures~\ref{plot:mloas-opt-action-dist-reapet1-1step-00observations} and \ref{plot:mloas-mrac-dist-reapet1-1step-00observations}, respectively.

With this in mind, we can see that, statistically, the performance of \textsc{RVerifyAC} and \textsc{DecPOMDP-OL} are equivalent since they select the same joint action and consider the same information in their performance.
We consider 2 variants of \textsc{\shortalgorithmname}, with different $\delta$ for the $\delta$-\NEPG strategy.
For $\delta\!=\!0.15$, the performance gap calculation in the $\delta$-\NEPG strategy is within the threshold, so no communication is triggered, but for $\delta\!=\!0.05$, the performance gap is beyond the threshold and so communication is triggered to improve the agent's performance.
Figure~\ref{plot:mloas-performance-gap-reapet1-1step-00observations} shows the affect of $\delta$ in this case.
The performance of \textsc{\shortalgorithmname-0.3-0.15} when executing the joint action (D+D), from the perspective to the agents, is \emph{worse} than the performance of \textsc{RVerifyAC-0.3} and \textsc{DecPOMDP-OL}.
This is since, according to the available information to the agents, (R+R) was the joint action with the highest objective in planning.
But from the centralized perspective, \textsc{\shortalgorithmname-0.3-0.15} is actually \emph{better} than \textsc{RVerifyAC-0.3} and \textsc{DecPOMDP-OL}, since it selects the same joint action as \textsc{MPOMDP-OL}.
In comparison, the performance of \textsc{\shortalgorithmname-0.3-0.05} when executing joint action (D+D), which includes communication between agents (due to $\delta$-\NEPG strategy), is \emph{better} than \textsc{RVerifyAC-0.3} and \textsc{DecPOMDP-OL} also from the agents perspective, bringing their performance closer to the performance of \textsc{MPOMDP-OL}, but with the cost of communication.

\begin{table*}[tb]
	\centering
	\begin{tabular}{|l|c|c|c|c|c|}

        \hline
        \multirow{2}{*}{Algorithm}  & \multirow{2}{*}{\# Inconsistencies}   & \multirow{2}{*}{\# Communications}    & \multicolumn{3}{c|}{ Final Return } \\
                                                                                                                    \cline{4-6}
                                    &                                       &                                       & Agent 1  & Agent 2  & Centralized \\
        \hline
        {\textsc{MPOMDP-OL}}  & 0.0\%  & 100.0\%  & \textbf{-7.44 $\pm$ 0.4}  & \textbf{-7.44 $\pm$ 0.4}  & \textbf{-7.44 $\pm$ 0.4} \\
        {\textsc{DecPOMDP-OL}}  & 50.0\% $\pm$ 25.0\%  & 0.0\%  & -7.95 $\pm$ 0.28  & -8.18 $\pm$ 0.34  & -7.57 $\pm$ 0.49 \\
        {\textsc{RVerifyAC}-0.8}  & 25.0\% $\pm$ 25.0\%  & \textbf{12.5\% $\pm$ 12.5\%}  & -7.89 $\pm$ 0.31  & -7.93 $\pm$ 0.57  & -7.55 $\pm$ 0.46 \\
        {\textsc{\shortalgorithmname}-0.8-0.1}  & 25.0\% $\pm$ 25.0\%  & \textbf{12.5\% $\pm$ 12.5\%}  & -7.89 $\pm$ 0.33  & -8.08 $\pm$ 0.36  & -7.45 $\pm$ 0.42 \\
        {\textsc{\shortalgorithmname}-0.8-0.05}  & 25.0\% $\pm$ 25.0\%  & 62.5\% $\pm$ 12.5\%  & \textbf{-7.76 $\pm$ 0.39}  & \textbf{-7.81 $\pm$ 0.4}  & \textbf{-7.42 $\pm$ 0.39} \\
        \hline
    \end{tabular}
	\caption{\small
        Statistical results over 75 runs with different unshared observations values and different seed, for 4 planning sessions in a 4x4 grid, with 2 unshared observations for each agent.
    }
	\label{tab:simulation-results-summary-4steps}
\end{table*}



Table~\ref{tab:simulation-results-summary-4steps} summaries statistical performance comparison in the 4x4 grid scenario with 4 planning sessions.
Algorithm \textsc{RVerifyAC}, with $\epsilon\!=\!0.8$, was able to reach better performance than \textsc{DecPOMDP-OL}, from the agents' perspective, with only {\char`~}12.5\% of the communications, while also reducing the amount of inconsistent action selections to {\char`~}25\%.
Yet, from the centralized perspective, the performance of \textsc{RVerifyAC-0.8} is about the same as the performance of \textsc{DecPOMDP-OL},
i.e., the agents (mostly) kept consistency, but still selected a suboptimal joint action with respect to the full joint history.
On the other hand, algorithm \textsc{\shortalgorithmname}, with $\epsilon\!=\!0.8$ and $\delta\!=\!0.1$,
also reduced the amount of inconsistent action selections to {\char`~}25\% with only {\char`~}12.5\% of the communications.
The performance of \textsc{\shortalgorithmname-0.8-0.1}, from the agent's perspective, is worse than the performance of \textsc{RVerifyAC-0.8}, but from the centralized perspective, \textsc{\shortalgorithmname-0.8-0.1} outperformed \textsc{RVerifyAC-0.8}, with performance approaching the \textsc{MPOMDP-OL}'s performance.
Additionally, for $\delta\!=\!0.05$, algorithm \textsc{\shortalgorithmname-0.8-0.05} was able to detect the gaps between the performance at planning and the performance at inference,
increasing the amount of communications to {\char`~}62.5\%, but improving the agents' performance.

While these results are demonstrated on relatively small-scale scenarios, we emphasize that the core advantages of our approach are general and can be extended to more complex and larger-scale problems.
Future work will focus on evaluating our method in more challenging environments.

\section{Conclusions}\label{sec:conclusions}

We addressed the challenge of decentralized multi-agent POMDP planning with limited data sharing capabilities. In such a setting, the agents' beliefs at planning time are inconsistent, in contrast to the typical assumption in Dec-POMDP approaches. Our  planning algorithm \algorithmname 
(i) features  a decentralized calculation of a consistent optimal joint action sequence by all agents with formal guarantees, where optimality is with respect to the full information of the system; and (ii) quantifies the probability distribution over the performance gap between planning with full information and inference with partial information available in practice to the agents. This distribution can then be used design strategies that trigger communication to improve the agent's performance at inference. We demonstrated the advantages of our approach in simulation,  considering a decentralized collaborative fire detection scenario, showing our algorithm outperforms state-of-the-art open-loop Dec-POMDPs planners. Future work includes extension to closed-loop planning, scaling to larger groups of agents, and further evaluation.

\bibliographystyle{plain}
\bibliography{refs}

\end{document}